\documentclass[pre,floats,aps,amssymb,preprint,epsf,epsfig,rotate]{revtex4}
\usepackage{graphicx}
\begin{document}


\draft

\title{Dynamic Critical approach to Self-Organized Criticality.}
\author{Karina Laneri, Alejandro F. Rozenfeld and Ezequiel V. Albano}

\affiliation{Instituto de Investigaciones Fisicoqu\'imicas Te\'oricas 
y Aplicadas, INIFTA. 
Sucursal 4, Casilla de Correo 16, (1900) La Plata, Argentina.}

\date{\today}


\begin{abstract}
A dynamic scaling Ansatz for the approach to the Self-Organized Critical (SOC) regime is 
proposed and tested by means of extensive simulations applied to the Bak-Sneppen model (BS),
which exhibits robust SOC behavior.     
Considering the short-time scaling behavior of the density of sites ($\rho(t)$) below the 
critical value, it is shown that i) starting the dynamics with configurations 
such that $\rho(t=0) \rightarrow 0$ one observes an {\it initial increase} of the
density with exponent $\theta = 0.12(2)$; ii) using 
initial configurations with $\rho(t=0) \rightarrow 1$, the density decays with 
exponent $\delta = 0.47(2)$.        
It is also shown that he temporal autocorrelation decays with exponent 
$C_a = 0.35(2)$. Using these, dynamically determined, critical exponents 
and suitable scaling relationships, all known exponents of the BS model can be obtained,
e.g. the dynamical exponent $z = 2.10(5)$, the  mass dimension exponent 
$D = 2.42(5)$, and the exponent of all returns of the activity $\tau_{ALL} = 0.39(2)$, 
in excellent agreement with values already accepted and obtained within the SOC regime.


\end{abstract}
\maketitle



Nearly two decades ago, Bak et al. \cite{pb} proposed the celebrated concept of 
Self-Organized Criticality (SOC) in order to describe complex systems capable of evolving toward 
a critical state without the need of tuning any control parameter. This is in contrast  to the 
case of standard critical behavior where critical points are reached by  tuning a suitable control 
parameter (temperature, pressure, etc.). The study of SOC behavior has attracted huge attention 
due to its ubiquity in a great variety of systems in the fields 
of biology (evolutionary models), 
geology  (earthquakes), physics (flick noise), zoology (prey-predators and herds), 
chemistry (chemical reactions), social sciences (collective behavior of individuals), 
ecology (forest-fire), neurology (neural networks), etc. \cite{book1,book2}.

In spite of the considerable effort invested in the study of SOC, 
it is surprising that little attention 
has been drawn to the understanding of the dynamic approach to the SOC regime when a system 
starts far from it. This issue is relevant for a comprehensive description of the phenomena since 
the SOC state behaves as an attractor of the dynamics.  For the case of
standard criticality, the existence of a {\bf short-time} universal dynamic scaling form
for model A  has only recently been established. In fact, according to a field-theoretical 
analysis followed by 
an $\epsilon-$expansion \cite{Schmittman},
which was subsequently been extensively confirmed by means of 
numerical simulations \cite{Huse},
a short-time universal dynamic evolution that sets in right after a time scale $t_{mic}$,
which is large enough in the {\it microscopic} sense but still very small in the
{\it macroscopic} one, has been identified. It is worth mentioning that by means of 
short-time measurement one can not only evaluate the dynamic exponent $z$ and 
relevant (static) exponents, but
also the exponent ($\theta$) describing the scaling behavior of the initial increase
of the order parameter of model A. 

Within this context, the aim of this work is to propose a dynamic scaling Ansatz to 
describe the approach to the SOC state. Furthermore, our proposal is validated by extensive
simulations of the evolutionary Bak-Sneppen  model, showing that the dynamic approach 
to the SOC state is in fact {\bf critical} and its study allows us to evaluate exponents
that are in excellent agreement with independent measurements performed within the SOC regime. 

{\it The Bak-Sneppen (BS) model} is aimed at simulating the evolution 
of life through individual mutations and their relation in the food 
chain \cite{bs,bs2,bs3,paczuski}. 
Each site of a $d-$dimensional array of side $L$ represents a species whose fitness is given 
by a random number $f$ taken from a uniform distribution $P(f)$ in the range $[0,1]$.
The system evolves according to the following rules:
(1) The site with the smallest fitness is chosen.
(2) A new fitness is assigned to that site, i.e., a random number taken from $P$. 
This rule is based on the Darwinian survival principle, i.e., the species with less 
fitness are replaced or mutated. 
(3) At the same time, the fitness of the nearest-neighbor sites are changed. 
This rule simulates the impact of the mutation over the environment.

The BS model is the archetype example of extremal dynamics 
and perhaps, it is the simplest example of a system exhibiting
a robust SOC behavior. We will perform  simulations in $d = 1$, assuming  
periodic boundary conditions. It is well known that the system reaches 
a stationary (SOC) state where the density $\rho$ of sites with 
fitness below a critical value is negligible 
($f < f_c$, with $f_c = 0.66702(3)$), but it is uniform 
above $f_c$ \cite{paczuski}.
Within the SOC state, the BS model 
exhibits scale-free evolutionary avalanches and punctuated 
equilibrium \cite{bs,paczuski}.  

{\it The dynamic scaling Ansatz.} The short-time dynamic scaling form has originally
been formulated for the Ising model with two state spins \cite{Huse}. In 
the absence of
magnetic fields, the Ising magnet exhibits a second-order phase transition when the 
temperature is tuned at the critical point (i.e., standard critical behavior).  
By analogy, we also considered that in the BS model, each site can only be 
in one of 
two possible states $\sigma_i$: occupied  ($\sigma_i = 1$)  when its fitness 
is below $f_c$ or empty  ($\sigma_i = 0$)   when its fitness is above $f_c$.
Furthermore, since the  magnetization goes to zero at the critical 
temperature of the 
Ising system, while in the BS model the density of sites $\rho$ also vanishes when 
the SOC regime is reached \cite{paczuski}, it is also reasonable to 
propose a scaling Ansatz 
for the density. However, it is worth mentioning that the BS model lacks a
control parameter, such as the temperature for the Ising model. Hence, in the 
limit $L\rightarrow\infty$, the proposed scaling reads
\vskip 2.0 true cm 
\begin{equation}
\rho (t,\rho_{0})=b^{-\aleph} \rho(b^{-z}t,b^{x_0}\rho_{0})
\label{scal-bs} 
\end{equation}
\noindent where $\rho_{0}$ is the initial density, $z$ is the dynamic exponent,
$x_0$ is the exponent of the rescaling of $\rho_{0}$,
$\aleph$ is an exponent, and $b$ is a scaling variable.
\vskip 0.75 true cm
\begin{figure}
\centerline{
\includegraphics[width=6cm,height=6cm,angle=0]{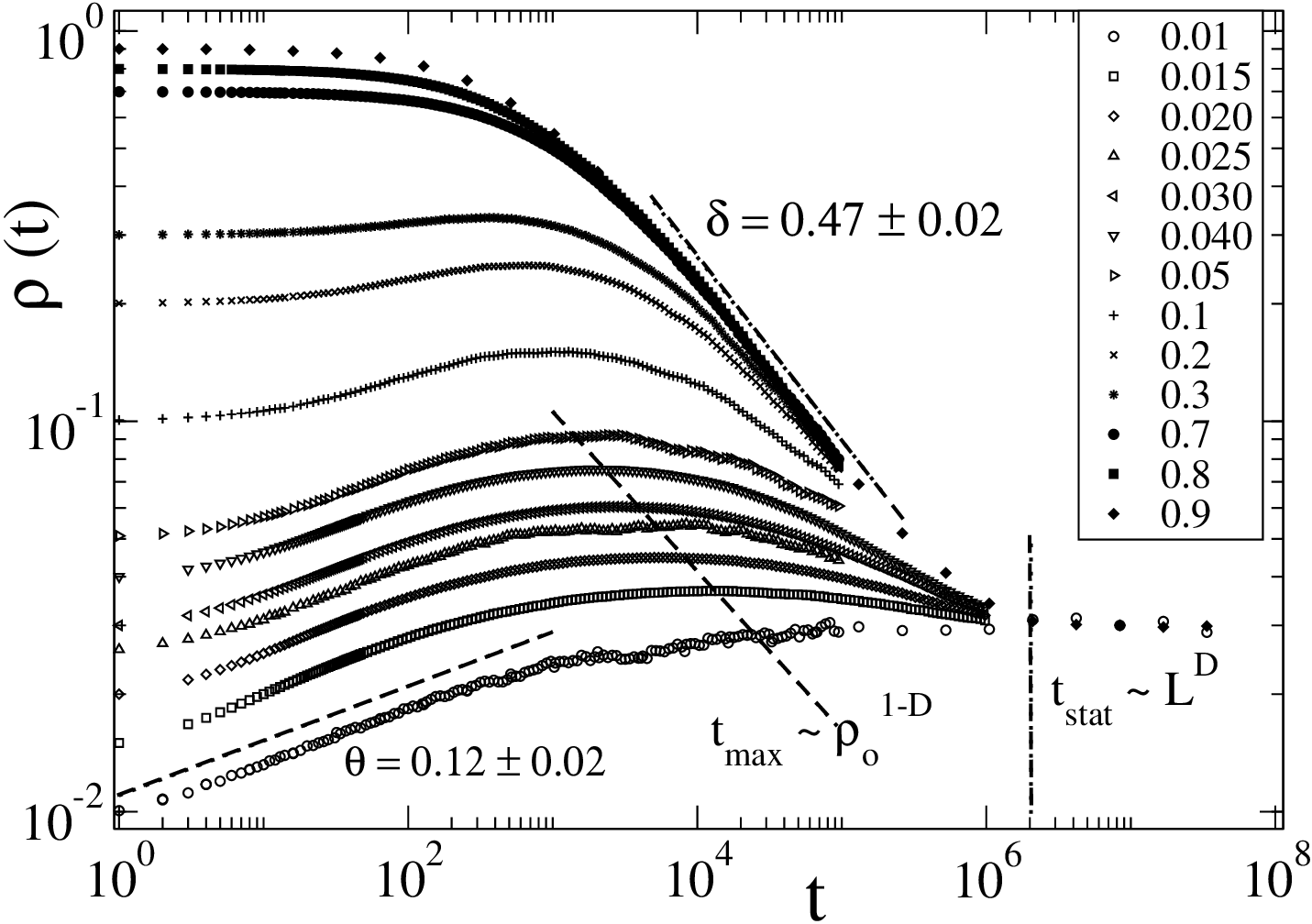}}
\caption{Temporal evolution of average density for several values of initial 
density $\rho_0$ as listed in the figure.}
\label{Fig1}
\end{figure} 
{\it Simulation results and discussion.} Computer simulations were performed in ensembles 
of $10^3$ different systems having the same initial density of sites with fitness 
below $f_c$. Notice that usually the site 
with the smallest fitness value at time $t$ is called {\it the active site}. 
Figure \ref{Fig1} shows the temporal evolution of density $\rho(t,\rho_0)$ as obtained 
for different values of $\rho_0$. Three different regimes can be distinguished: 
i) A {\bf short-time regime} $(0<t<t_{max} \approx 10^{3})$, which holds for 
low initial densities $\rho_{0} < 0.1$, where the density exhibits an 
initial increase. Since the value of the ``effective'' exponent 
slightly depends on $\rho_{0}$, as ussually \cite{Huse}, 
we have performed an extrapolation 
obtaining $\theta = 0.12(2)$ in the $\rho_{0} \rightarrow 0$ limit.
ii) An {\bf intermediate-time regime} $(t_{max}<t<t_{stat} \approx 10^{6})$ 
where the density decreases also following a power-law with exponent $\delta=0.47(2)$,
and iii) A {\bf long-time regime} $(t_{stat} < t)$, where the 
system arrives at a stationary state with a constant average density $\rho_{stat}$.

In order to understand the observed behavior it is useful to analyze first 
a simulation started with a single site ($\rho_0=1/L$).
One observes (not shown here for the sake of space) a single avalanche and the 
density increases monotonically with exponent $\theta$, as in the case of 
figure \ref{Fig1} for $\rho_0 \rightarrow 0$. The spatiotemporal evolution 
of the avalanche is delimited according to 

\begin{equation}
r(t) \sim t^{1/D} ,
\label{ec:crecimiento}
\end{equation}

\noindent where $D=2.43(1)$ \cite{paczuski} is {\it the mass dimension} exponent.

Then it is convenient to consider the evolution of $n_e$ epidemics started simultaneously 
with a separation of $r_e$ empty sites between them. Two neighbor epidemics collide
when each of them expands its activity over $\frac{r_e}{2}$ sites,
on average. Then, the border between them disappears leading to 
a single epidemic. So, according to equation (\ref{ec:crecimiento}), this requires a 
time of order $t_e \sim (r_e/2)^D$, hence for the collision of $n_e$ epidemics the 
time required is of order $t \sim n_e t_e$. Then, one expects an initial increase
of the density until $t_{max}$, given by
\begin{equation}
t_{max}\sim n_e \left(\frac{r_e}{2}\right)^D .
\label{ec:t_max}
\end{equation}
Now, for an initial random distribution, one has that $n_e = L \rho_{0}$ and 
the distance between particles at $t = 0$ 
is of order $r_e \sim 1/\rho_0$. Then, equation (\ref{ec:t_max}) becomes
\begin{equation}
t_{max}\sim \rho_0^{1-D} .
\label{cabronzuelos}
\end{equation}
Also, the system reaches the stationary  state for a time of order 
$t_{stat} \sim L^D $ \cite{paczuski}. In the thermodynamic limit $\rho_{stat}\rightarrow 0$ 
and $t_{stat}\rightarrow \infty$.

It is worth mentioning that the time $t$ used in this work   
is a discrete sequential time. An alternative definition corresponds to 
the {\it parallel time} $t_{\parallel}$, usually employed to 
define the dynamic exponent $z$ according to  
\begin{equation}
t_{\parallel}\sim r^z.
\label{ec:exp_dinamico}
\end{equation}
The unit of parallel time is defined as the average number of actualization 
steps that have to be performed to change the state of all the occupied 
sites ($n(t)$), then
\begin{equation}
t \rightarrow t+1 \qquad  \text{and} \qquad
t_{\parallel} \rightarrow t_{\parallel}+\frac{1}{n(t)}
\label{hdp}
\end{equation}
\noindent From these definitions one has $t_{\parallel}\sim t^{1-\theta}$,
which inserted in equations (\ref{ec:exp_dinamico}) 
and (\ref{ec:crecimiento}), yields 
\begin{equation}
1-\theta=\frac{z}{D} .
\label{ec:D_z}
\end{equation}

The {\it time scaling behavior} of the density can be obtained from  equation (\ref{scal-bs})
by replacing $b = t^{1/z}$, which yields
\begin{equation}
\rho (t,\rho_{0})=t^{-\aleph/z} \Phi(t^{x_0/z}\rho_{0})
\label{scal-bs-t}
\end{equation}
\noindent where $\Phi$ is a scaling function. For  $\rho_0 \rightarrow 1$ 
and within the intermediate time regime, $\rho(t)$ becomes independent 
of $\rho_0$ (see figure \ref{Fig1}), thus assuming $x_0/z>0$ one has
$\Phi(t^{x_0/z}\rho_{0} \gg 1) = constant$, which holds for 
$t \gg t_{max} \propto \rho_{0}^{-z/x_{0}}$. Then, $t_{max}$ sets the 
time scale for the initial increase of the density, as shown in 
figure \ref{Fig1} \cite{Note}. Furthermore, according to 
equation (\ref{cabronzuelos}) the following relationship between exponents 
should hold 
\begin{equation}
D - 1 = z/x_{0} .
\label{cachondos}
\end{equation}
Also, $\rho$ should decrease according to
\begin{equation}
\rho(t) \sim t^{-\delta} ,
\label{ec:alef-z}
\end{equation}
\noindent with $\delta \equiv \aleph/z = 0.47(2)$.
On the other hand, for $\rho_{0} \rightarrow 0 $ and within the short-time regime 
($t \ll t_{max}, t \rightarrow 0 $) the dependence of $\Phi$ on the 
initial density becomes relevant, thus we assume $\Phi(x)\sim x^u$,
where $u$ is an exponent. Hence, replacing in equation (\ref{scal-bs-t}), one has
\begin{equation}
\rho (t,\rho_{0}) = \rho_{0}^u t^ \theta ,
\label{ec:teta}
\end{equation} 
\noindent where $\theta = -\delta + u \frac{x_0}{z}$. So, $\theta = 0.12(2)$
is the exponent that describes the initial increase of the density within the
short-time dynamic critical regime of the BS model.  

\begin{figure}
\centerline{
\includegraphics[width=7cm,height=5cm,angle=0]{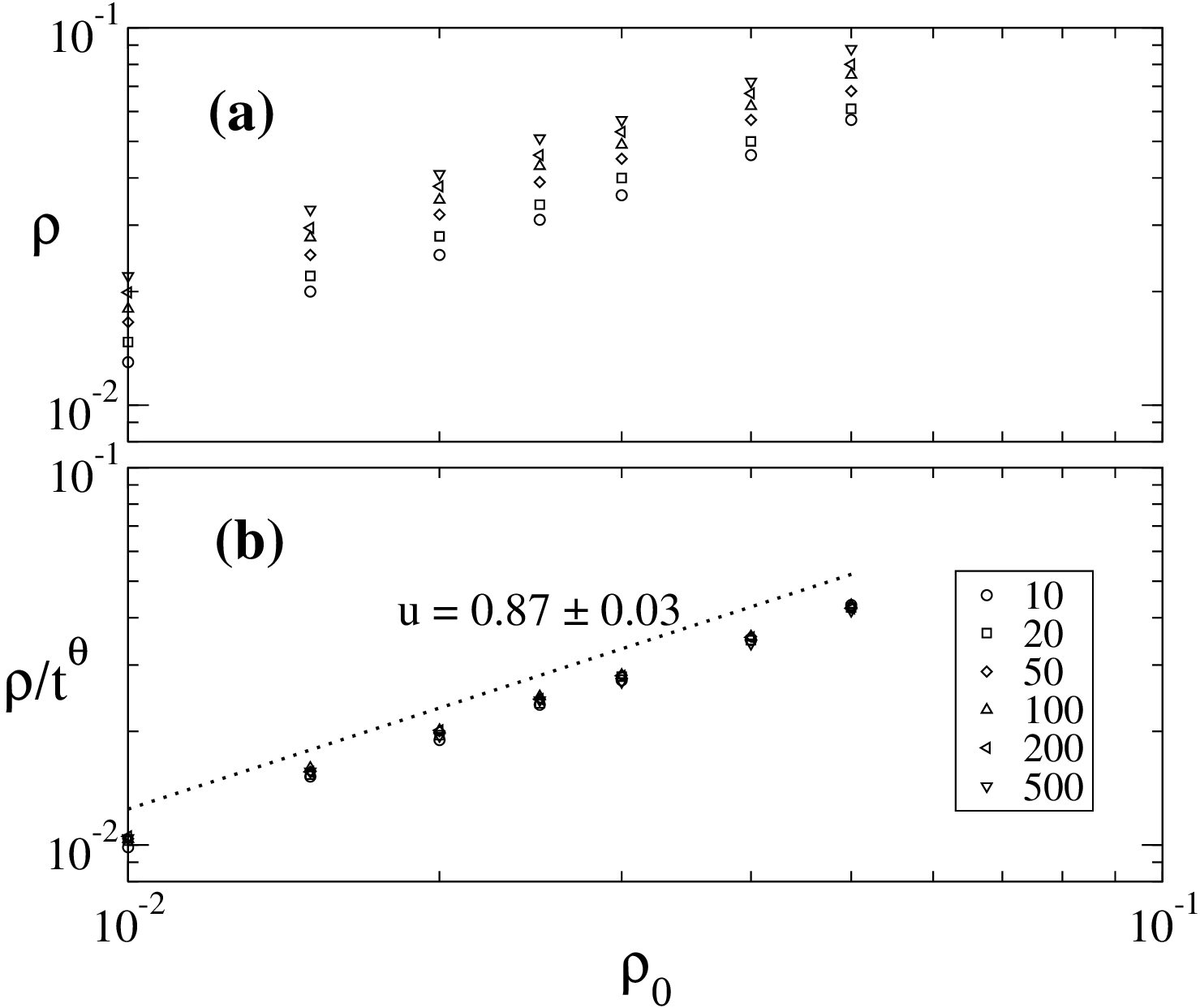}}
\caption{(a) Log-log plots of $\rho$ versus $\rho_0$ obtained for 
times $t = 10, 20, 50, 100, 200$, and $500$ (from bottom to top).
(b) Scaling plot of the curves shown in (a) using equation (\ref{ec:teta}).}
\label{Fig5}
\end{figure}

In order to calculate $u$, the dependence of the density 
on $\rho_{0}$ (i.e., $\rho(\rho_{0})$) was measured 
at different times, as shown in figure \ref{Fig5} (a). 
Also, the scaling Ansatz suggested by 
equation (\ref{ec:teta}) is shown in figure \ref{Fig5} (b). The observed 
data collapse is  satisfactory and the exponent $u = 0.87(3)$ was measured. 
  
\begin{figure}
\centerline{
\includegraphics[width=7cm,height=5cm,angle=0]{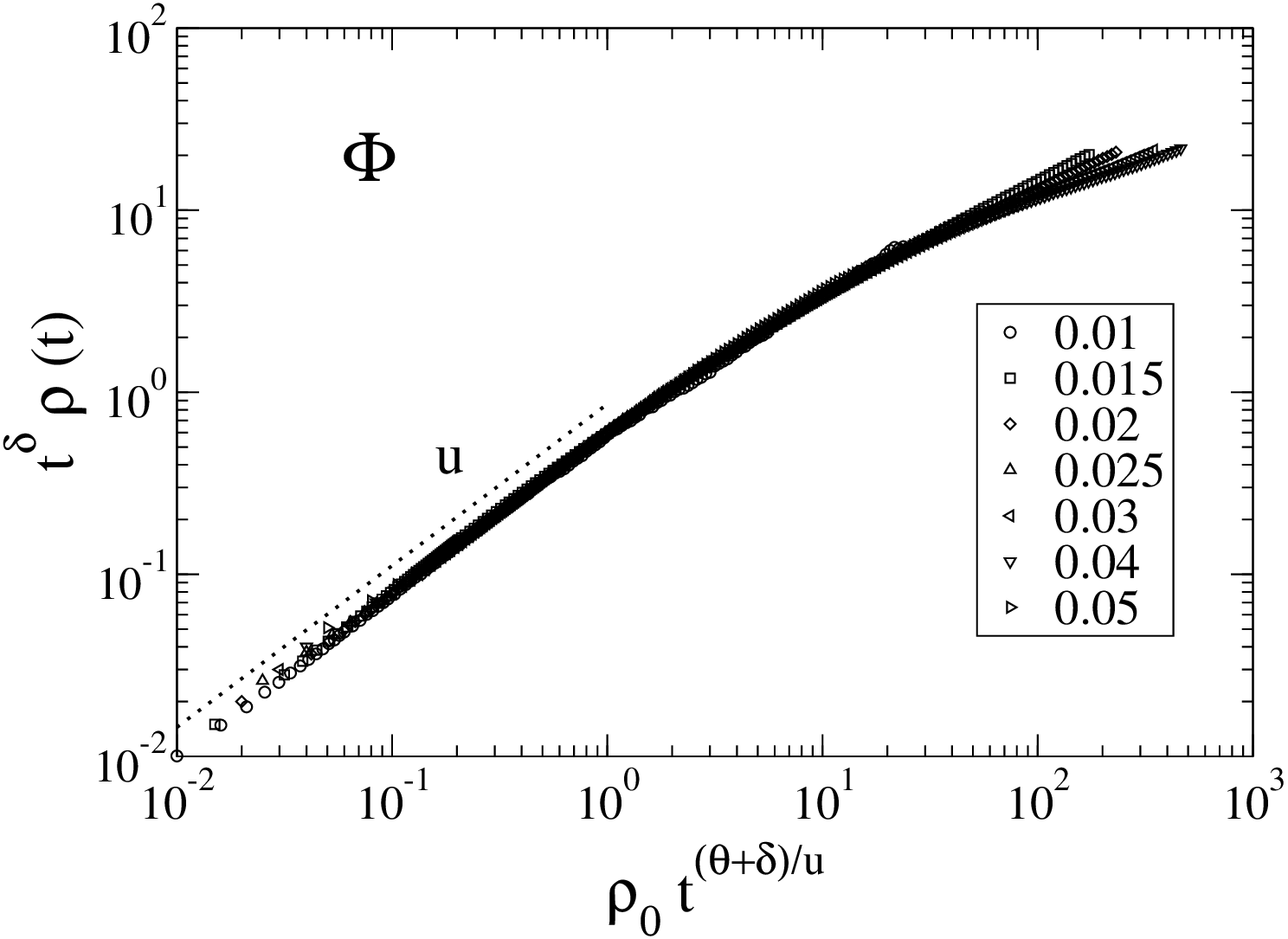}}
\caption{Scaling plot of the data shown in figure \ref{Fig1} obtained by 
using equation (\ref{ec:fi}) for the time-scaling behavior.
The dotted line -slightly shifted up for the sake of clarity- 
with slope $u = 0.89$ corresponds to the best fit
of the data obtained for the lowest density.}
\vskip 0.50 true cm
\label{Fig6}
\end{figure}

On the other hand, replacing $\frac{x_0}{z} = \frac{\theta+\delta}{u} $
in equation (\ref{scal-bs-t}) one obtains

\begin{equation}
\rho (t,\rho_{0})=t^{-\delta}\Phi(\rho_{0} t^{\frac{\theta+\delta}{u}}) .
\label{ec:fi}
\end{equation}
The shape of the scaling function $\Phi$ is shown in figure \ref{Fig6} and the  
excellent data collapse obtained by plotting the data already shown 
in figure \ref{Fig1}, strongly supports the formulated scaling hypothesis 
and the calculated exponents. In fact, the best fit of the 
data obtained, for $\rho_{0} \rightarrow 0$, gives $u = 0.89(3)$ (see dotted line
covering two decades in figure \ref{Fig6}) that is in agreement with the 
preliminary estimation
already performed with the data shown in figure \ref{Fig5}
(covering less than one decade).

Also, {\it the scaling of the initial density} can also be obtained by
replacing $b=\rho_0^{-1/x_{0}}$ in equation (\ref{scal-bs}), giving 

\begin{equation}
\rho (t,\rho_{0})=\rho_{0}^{\aleph/x_0} \Omega(t\rho_{0}^{z/x_0}) ,
\label{ec:omega}
\end{equation}
\noindent where $\Omega$ is a scaling function that, for $t \rightarrow 0$ and
$\rho_{0} \rightarrow 0$, can be approximated by $\Omega(x)\rightarrow x^v$ 
($x \rightarrow 0$), where $v$ is an exponent. Thus one obtains

\begin{equation}
\rho (t,\rho_{0})=\rho_{0}^{\aleph/x_0+v \frac{z}{x_0}} t^v .
\label{scal-bs-rho}
\end{equation}

\noindent Now, by comparing equation (\ref{scal-bs-rho}) with equation (\ref{ec:teta})
it follows that $v = \theta$ and $u=\frac{\aleph}{x_0}+v\frac{z}{x_0}$. Hence, 

\begin{equation}
\frac{\aleph}{x_0}=\frac{\aleph/z}{x_0/z}=\frac{u\delta}{\delta+\theta}
\label{ec:alef_x0}
\end{equation}

Then, using $\frac{x_0}{z} = \frac{\theta+\delta}{u} $ and equation (\ref{ec:alef_x0}),
one has that equation (\ref{ec:omega}) can be written in terms of already measured 
exponents, so 

\begin{equation}
\rho (t,\rho_{0}) = 
\rho_{0}^{\frac{u\delta}{\delta+\theta}} \Omega(t\rho_{0}^{\frac{u}{\delta+\theta}}).
\label{omega1}
\end{equation}

\begin{figure}
\centerline{
\includegraphics[width=7cm,height=5cm,angle=0]{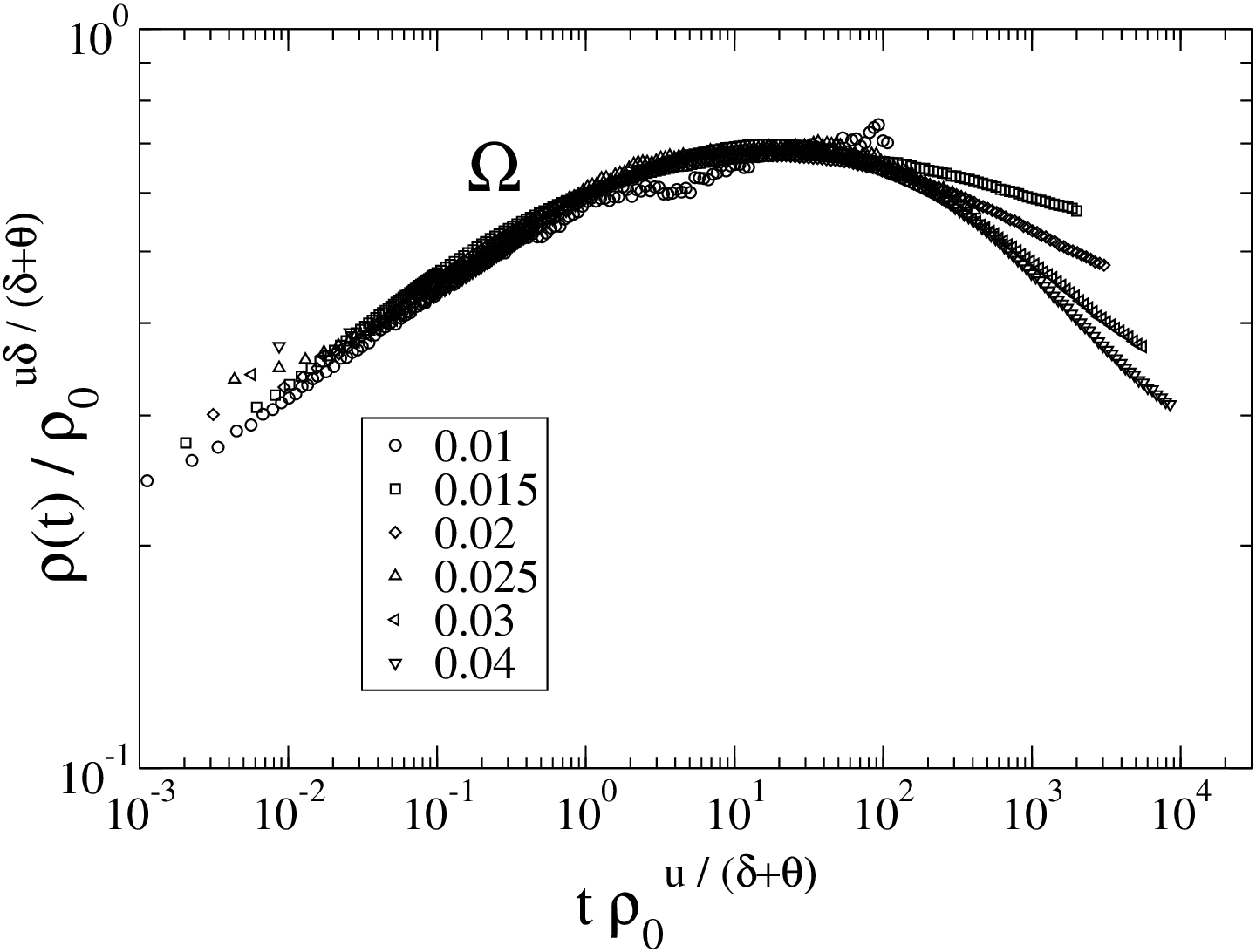}}
\caption{Scaling plot of the data shown in figure \ref{Fig1}
obtained by using equation (\ref{omega1}) for the scaling of the initial density.}
\label{Fig7}
\end{figure}

Figure \ref{Fig7} shows plots of the numerical data performed according to 
equation (\ref{omega1}). The shape of the scaling function $\Omega$ can be observed and 
the collapse of the curves also supports the formulated scaling hypothesis.

Valuable information on the dynamic behavior of the system can also be obtained
by measuring the {\it temporal autocorrelation of the state of the site} 
($A(t,t_0)$, averaged over all sites)
that is expected to decay according to a power law \cite{Schmittman,Huse}, namely

\begin{eqnarray}
A(t_0,t) = \langle \sigma_i(t_0) \sigma_i(t) \rangle - 
\langle \sigma_i(t_0) \rangle   \langle \sigma_i(t) \rangle
\sim t^{-C_a} 
\label{ec:A-pl} 
\end{eqnarray}

\noindent where $\sigma_{i}(t) = 1,0$ is the state of the site at time $t$.
Also \cite{Huse}

\begin{eqnarray}
C_a=d/z-\theta. 
\label{ec:A-Ca}
\end{eqnarray}

\begin{figure}
\centerline{
\includegraphics[width=7cm,height=5cm,angle=0]{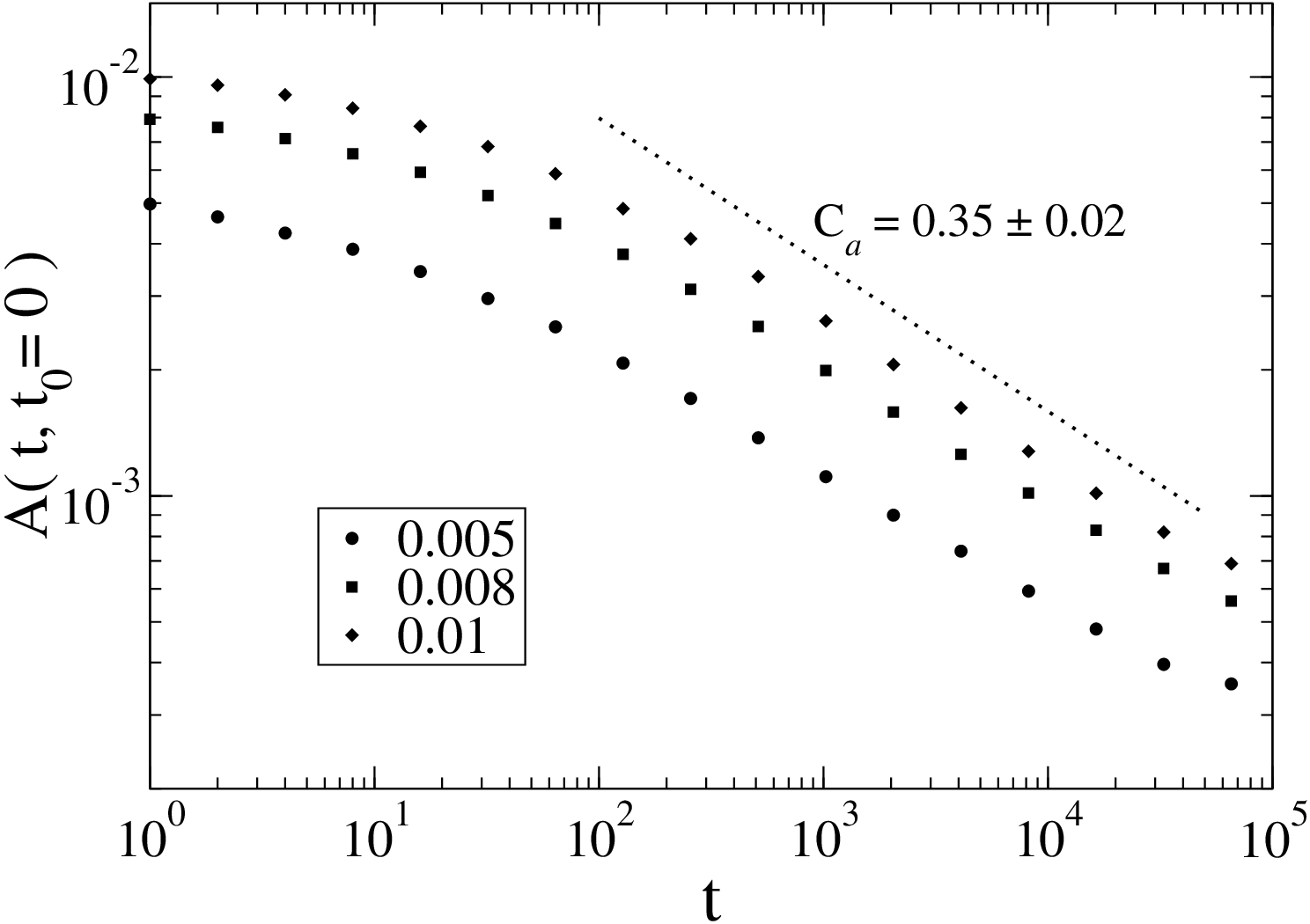}}
\caption{Log-log plots of the temporal autocorrelation versus $t$ obtained for
different values of $\rho_0$. The dotted line with slope $C_{a} = 0.35$ has 
been drawn for the sake of comparison. }
\label{Fig8}
\end{figure}

Figure \ref{Fig8} shows log-log plots of the autocorrelation versus $t$ obtained 
using very low initial densities since equation (\ref{ec:A-pl}) is expected to hold
for $\rho_{0} \rightarrow 0$. From these plots one determines $C_a = 0.35(2)$,
and replacing this value in equation (\ref{ec:A-Ca}) the dynamic exponent 
$z = 2.13(5)$ is obtained. Then by using equation (\ref{ec:D_z}) 
we obtain $D=2.42(5)$. These exponent values are in agreement with 
those already published in the literature that were obtained within the SOC regime
\cite{paczuski}, i.e., $z=2.1(5)$  and $D = 2.43(1)$. 
Furthermore, equation (\ref{cachondos}) can be tested, yielding 
$D -1 = 1.42(5)$ (left-hand side) 
and $z/x_{0} = u/(\theta + \delta) = 1.47(5)$ (right-hand side).

Now, we can discuss the relationship between short-time dynamic measurements 
and the behavior of relevant distributions characterizing return times of the 
activity to a given point in space. In fact, the distribution 
$P_{ALL}(0,t_{\parallel})$ 
is the probability for the activity at time $t_{\parallel}$ 
to revisit a site that was visited at time $0$ and is often referred 
to as the distribution of all return
times \cite{paczuski}. For $t_{\parallel} \gg 1$ such 
a distribution decays as a power law
of the form $P_{ALL}(0,t_{\parallel}) \propto t_{\parallel}^{-\tau_{ALL}}$, 
where $\tau_{ALL}$ is the 
'lifetime' exponent for all returns of the activity \cite{paczuski}. Then, 
considering that $A(0,t)$ is the probability that a given site 
being occupied at 
$t =0$ returns to be occupied at time $t$, one has that 
$A(0,t) \propto P_{ALL}(0,t_{\parallel})$ since a site can 
change its state only when such a site, or any of its two neighbours, 
are visited by the activity.   
Then, by recalling that we are working in terms of the 
discrete sequential time $t$ such as 
$t_{\parallel} \propto t^{1-\theta}$ \cite{note1}, one has 

\begin{equation}
A(t,0)\sim t^{-(1 - \theta)\tau_{ALL}}, \qquad  \text{with} \qquad
\tau_{ALL} \equiv \frac{C_a}{1-\theta} .
\end{equation}

By means of short-time dynamic measurements we have obtained 
$\tau_{ALL} = 0.39(4)$, in excellent agreement with the value 
calculated within the stationary state ($t>10^6$), 
given by $\tau_{ALL} = 0.42(2)$ \cite{paczuski}.

Summing up, the proposed dynamic scaling Ansatz for the approach to the 
stationary state generalizes the concepts previously developed to 
describe the critical dynamics of model A for systems exhibiting SOC. 
The dynamic scaling behavior was tested with the BS model showing that it 
holds for the density of sites with fitness below the critical one. 
Remarkably the exponents calculated by using the dynamic approach are in 
excellent agreement with those already measured within the stationary state.
Using well established relationships between exponents and considering 
that only two of them are basic, the dynamic measurement allows the  
self-consistent  evaluation of all exponents of the BS model.
So, we have shown that the dynamic approach to the SOC regime is indeed critical and 
it is governed by the dynamic exponent $z$, standard exponents 
(i.e., those available from stationary determinations), and by the exponent
$\theta$ that is introduced to describe the initial increase of the density. 
It is worth  emphasizing that the criticality observed in the approach to the SOC state
not only addresses new and challenging theoretical aspects of SOC behavior, but also  
is of great practical importance for the determination of relevant exponents.

\vskip 1.0 true cm


\begin{thebibliography}{99}
\bibitem{Note} Notice that in magnetic systems the time scale of the initial
increase of the magnetization is set by $t_{max} \propto m_{o}^{-z/x_{0}}$,
where $m_{0} \rightarrow 0$ is the initial magnetization \cite{Huse}.
\bibitem{note1} The relevant time scales involved are discussed in connection to
equations (\ref{ec:exp_dinamico}) and (\ref{hdp}).
\end{thebibliography}
\end{document}